\documentclass[11pt]{article}
\usepackage[margin=1in]{geometry}
\usepackage{times}
\usepackage{microtype}
\usepackage{setspace}
\usepackage{booktabs}
\usepackage{tabularx}
\usepackage{array}
\usepackage{tikz}
\usetikzlibrary{positioning,arrows.meta}
\usepackage{graphicx}
\usepackage{xcolor}
\usepackage{titlesec}
\usepackage{hyperref}
\hypersetup{
  colorlinks=true,
  linkcolor=black,
  citecolor=black,
  urlcolor=blue!55!black,
  breaklinks=true
}
\usepackage{enumitem}
\setlist{nosep,leftmargin=1.5em,topsep=2pt,itemsep=1.5pt}
\titlespacing*{\section}{0pt}{10pt plus 2pt minus 2pt}{4pt plus 1pt minus 1pt}
\titlespacing*{\subsection}{0pt}{8pt plus 2pt minus 2pt}{3pt plus 1pt minus 1pt}
\titlespacing*{\subsubsection}{0pt}{5pt plus 1pt minus 1pt}{2pt plus 1pt minus 1pt}

\newcommand{\code}[1]{\texttt{\small #1}}
\newcommand{\ghrepo}{https://github.com/dynaroars/vietprofs}
\newcommand{\codelink}[1]{\href{\ghrepo/blob/main/#1}{\code{#1}}}
\newcommand{\codelinkd}[2]{\href{\ghrepo/blob/main/#1}{\code{#2}}}
\newcommand{\commitlink}[1]{\href{\ghrepo/commit/#1}{\code{#1}}}
\newcommand{\vplink}[2]{\href{https://vietprofs.roars.dev/?q=#1}{#2}}

\title{\includegraphics[width=0.13\textwidth]{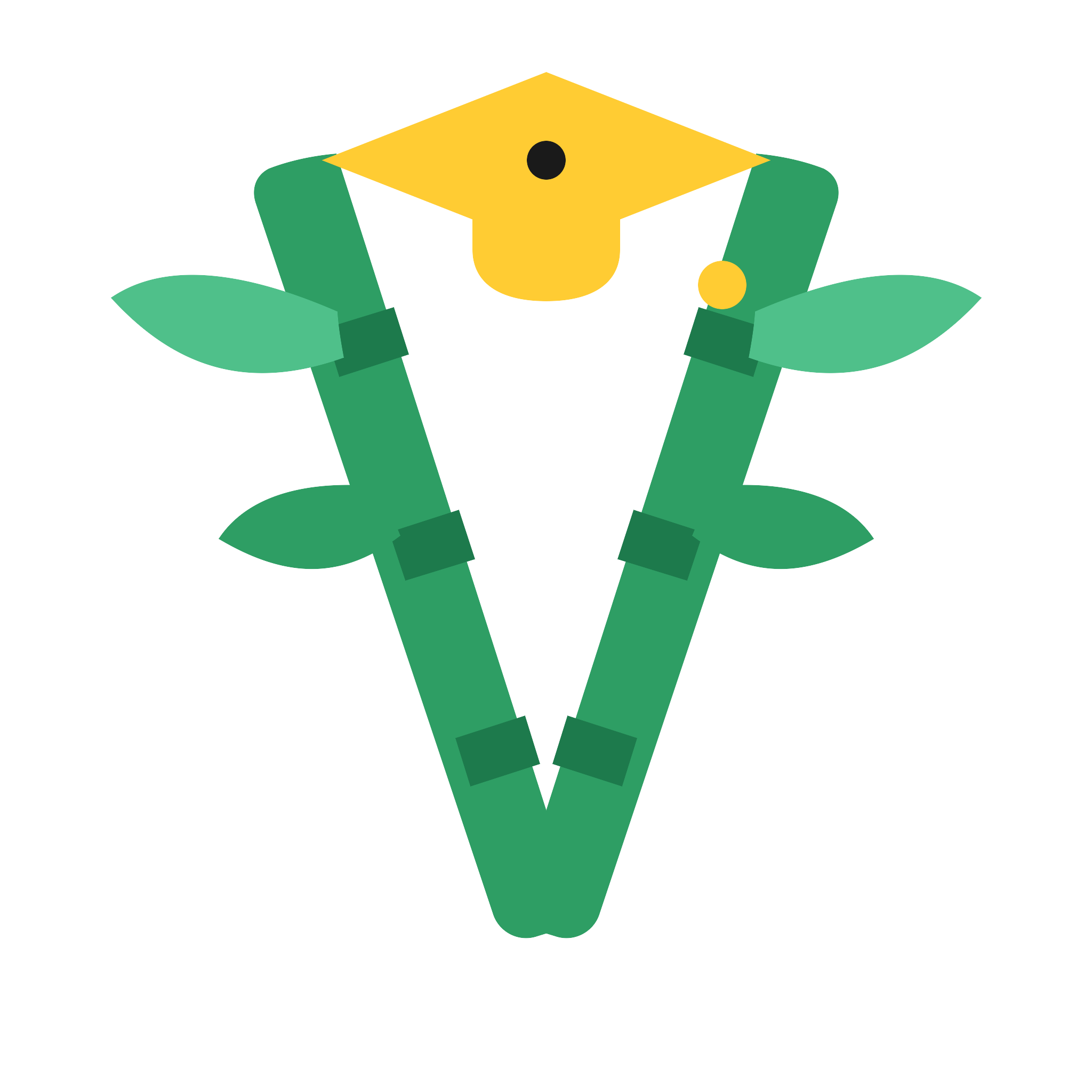}\\[-2pt]
VietProfs: A Public Directory of the Vietnamese Academic and Research Diaspora\\[4pt]
\large Purpose, Eligibility, and Design, with a Technical Report on Keeping It Accurate}
\author{ThanhVu Nguyen\\George Mason University\\}
\date{September 2026}

\begin{document}
\maketitle

\begin{abstract}
The Vietnamese academic diaspora spans hundreds of universities and research institutes worldwide, yet there has never been a centralized, searchable directory to find scholars in this community.
Prospective graduate students looking for advisors who share their background, researchers seeking collaborators, and organizers looking for speakers have had to rely on manual, one-off web searches.
To address this need, we built \textbf{VietProfs} (\url{https://vietprofs.roars.dev}), an open, searchable directory of Vietnamese and Vietnamese-diaspora faculty and permanent research scientists outside Vietnam.
Each profile pairs a portrait and authoritative diacritic name with verified appointment, degree, and honors data, browsable through a fast client-side web interface.

This paper covers two topics: the design of the directory itself and the engineering required to keep it current.
Part~I describes the directory: its goals, our specific eligibility and appointment track rules, how its 17 fields align with standard U.S. federal classifications (NCES, NSF, and NIH), the design of individual profile cards, and what the current roster shows.
Part~II provides a technical report on automated directory maintenance.
Keeping over a thousand records accurate against a constantly shifting web is notoriously difficult.
We describe an LLM-assisted maintenance system that combines automated search and periodic revalidation with deterministic validation gates.
Our core operating rule is simple: \emph{AI proposes, evidence decides}.

As of September 2026, the directory includes 1,152 verified records across 492 institutions in 23 countries, covering doctoral cohorts from 1939 to 2026.
Aggregate statistics are computed dynamically at runtime rather than stored statically, and all changes are audited through public Git history.
VietProfs provides a practical, long-overdue community resource while serving as a concrete case study in building reliable, agent-maintained public datasets.
\end{abstract}

\section{Introduction}
\label{sec:intro}
Vietnamese and Vietnamese-diaspora faculty and research scientists work across hundreds of universities and research institutes worldwide.
Despite the size and scientific contributions of this population, there has never been a centralized, searchable record of it.
A student looking for a graduate advisor who shares their cultural background, a faculty member looking for compatriots at nearby institutions, or an organizer assembling a workshop panel has had no better option than ad hoc Google searches, one name at a time.
In other diaspora communities, informal rosters maintained by alumni networks or professional associations have long played an important role in fostering mentorship, scientific collaboration, and knowledge sharing back toward the country of origin \cite{meyer1999scientific,kerr2008ethnic}.

We created \textbf{VietProfs} to fill this gap: an open, searchable directory of Vietnamese and Vietnamese-diaspora faculty and faculty-equivalent research leaders.
However, building a directory is only the first step; the harder engineering challenge is keeping it accurate over time as people move, get promoted, retire, or change institutions.
By combining AI agents with strict deterministic verification against the public web, the open-source system maintains itself with minimal manual overhead, preventing the directory from becoming obsolete or abandoned.

\subsection{What VietProfs Is: Motivation, Goals, and Design}
\label{sec:intro-what}
Before this project, Vietnamese-heritage scholars could only be discovered piecemeal across scattered departmental websites, with no way to see the community as a whole.
VietProfs brings these scholars together into a single, cohesive directory.
The site serves several concrete practical uses:
\begin{itemize}
\item \textbf{Finding a mentor or advisor:} Helping prospective students find graduate advisors and mentors with shared backgrounds and research interests.
\item \textbf{Connecting across institutions:} Enabling faculty to discover colleagues across institutions and geographical regions.
\item \textbf{Organizing events:} Assisting conference organizers, professional societies, and student associations in identifying speakers, panelists, and committee members.
\item \textbf{Documenting achievements:} Providing journalists, academic institutions, and cultural organizations with a reliable starting point for documenting diaspora contributions.
\item \textbf{A shared record for the community:} Giving the Vietnamese diaspora a shared record of its own reach, spanning from early pioneers like \vplink{Buu-Hoi+Nguyen-Phuc}{Buu-Hoi Nguyen-Phuc} (1939 Paris doctorate; CNRS Director of Research) and \vplink{Xuong+Nguyen-Huu}{Xuong Nguyen-Huu} (1962 UC Berkeley PhD; UC San Diego) to recent assistant professors (Section~\ref{sec:goals}).
\end{itemize}

Inclusion is governed by a clear, name-independent rule: a verified academic appointment at a university or eligible public/nonprofit research institute outside Vietnam, in one of seven accepted tracks---Tenure-line, Teaching, Research, Clinical, Academic staff, Emeritus, or Deceased (Section~\ref{sec:eligibility}).
The Research track includes permanent, faculty-equivalent researchers at public bodies like CNRS, INRIA, Max Planck, RIKEN, and CSIRO, who lead groups, secure grants, and mentor students much like university professors.
Temporary, visiting, postdoctoral, and adjunct roles are excluded to keep the directory stable.

Each scholar is mapped to one of 17 broad fields of study (Section~\ref{sec:fields}), adapted from classifications used by the U.S. National Center for Education Statistics (NCES), National Science Foundation (NSF), and National Institutes of Health (NIH).
This ensures our field distribution corresponds to familiar national categories rather than an arbitrary schema.

As of September 2026, the directory contains 1,152 verified records across 492 institutions in 23 countries (Section~\ref{sec:results}).
Each profile card is designed around what visitors actually need: a portrait, authoritative Vietnamese-diacritic name, current university, department, rank, location, educational background, curated honors, and research tags (Section~\ref{sec:whatshown}).
The web interface supports instant text search with optional single-field scoping, combinable filters, shareable URLs, dynamic leaderboards, a U.S. state map, and a public submission form for proposed additions and corrections (Section~\ref{sec:ui}).

\subsection{The Engineering Challenge: Keeping VietProfs Accurate}
\label{sec:intro-how}
Compiling an initial directory is straightforward; keeping it accurate over time is the real engineering problem, and it forms the technical focus of this paper.
Public directories decay quickly because the world changes while databases remain static: faculty change universities, get promoted, transition to emeritus status, or retire.
Furthermore, institutional web pages frequently redesign or break links \cite{bar-yossef2004sic}.
Manually auditing hundreds of profiles against the live web is time-consuming, tedious, and unscalable.

VietProfs addresses this challenge with an automated maintenance system based on two continuous loops: \emph{expansion} (discovering eligible scholars not yet in the roster) and \emph{revalidation} (regularly checking existing records against the live web to refresh appointments, ranks, degrees, and links; Section~\ref{sec:maintenance}).
These automated workflows run continuously to ensure the directory stays fresh.

One practical lesson from building this system is the fundamental difference between candidate discovery and claim verification.
While modern LLMs are effective at extracting structured facts from a given webpage, they struggle with open-ended, unconstrained searches for new scholars.
In our experience, letting agents blindly crawl university websites burns substantial time and API tokens with diminishing returns, often rediscovering existing scholars or returning ineligible postdocs.
In contrast, crowdsourced leads from the community via our public submission form (Section~\ref{sec:submit}) provide an excellent source of new candidates.
The agent pipeline can then do what it does best: verify the submitted claims against official university pages.
Because LLMs can still make errors---such as matching a photo to the wrong person---human review remains a crucial backstop for high-impact edits (Section~\ref{sec:architecture}).

This paper is structured in two parts:
\begin{itemize}
\item \textbf{Part~I} presents VietProfs as a public resource: its purpose, eligibility criteria, disciplinary classifications, displayed fields, and current snapshot.
\item \textbf{Part~II} details the engineering system: our data model and architecture (Section~\ref{sec:model} onward), the discovery and verification pipeline (Sections~\ref{sec:architecture}--\ref{sec:evidence}), the revalidation engine (Section~\ref{sec:maintenance}), the testing and safety gates (Section~\ref{sec:testing}), and practical engineering lessons learned from running the system in production (Sections~\ref{sec:whatdoesnotwork}--\ref{sec:conclusion}).
\end{itemize}

\begin{center}
\setlength{\fboxsep}{6pt}
\setlength{\fboxrule}{0.5pt}
\fbox{\parbox{0.55\linewidth}{\centering
\textsc{VietProfs is live and searchable at}\\[2pt]
{\Large\url{https://vietprofs.roars.dev}}
}}
\end{center}

\vspace{0.6em}
\noindent\rule{\linewidth}{0.4pt}
\begin{center}
{\Large\textbf{Part I: VietProfs: Purpose, People, and Content}}
\end{center}
\noindent\rule{\linewidth}{0.4pt}
\vspace{0.4em}

\section{Goals and Intended Uses}
\label{sec:goals}
VietProfs addresses a straightforward need: if someone wants to find Vietnamese-heritage scholars in a specific country, university, or field, where can they look?
Until now, there was no centralized answer.
We designed VietProfs around several primary use cases:

\begin{itemize}
\item \textbf{Finding a mentor or advisor.} Prospective graduate students seeking advisors who share their background, or who want Vietnamese-speaking mentors, can filter by field and location rather than guessing names from university department rosters one by one. For example, a search for Engineering quickly surfaces established researchers like \vplink{Nam-Trung+Nguyen}{Nam-Trung Nguyen} at Griffith University or \vplink{Tho+Le-Ngoc}{Tho Le-Ngoc} at McGill University.
\item \textbf{Connecting across institutions.} A faculty member moving to a new university, or looking for colleagues in the same metropolitan area, can search by city or country to connect with peers. For instance, a researcher in Singapore can readily identify colleagues such as \vplink{Nhan+Phan-Thien}{Nhan Phan-Thien}, \vplink{Hai+Minh+Duong}{Hai Minh Duong}, and \vplink{Han+Vinh+Huynh}{Han Vinh Huynh} at the National University of Singapore.
\item \textbf{Organizing conferences and committees.} Organizers of academic conferences, workshops, and student associations can use field and honors filters to find keynote speakers, session chairs, and committee members with specific expertise.
\item \textbf{Documenting diaspora contributions.} Journalists, university diversity offices, and cultural organizations now have an authoritative, verified directory to consult when highlighting Vietnamese scholarly achievements abroad.
\item \textbf{A shared record for the community.} Beyond individual queries, the directory offers a comprehensive view of the diaspora's intellectual footprint, documenting contributions across multiple generations---from 1939 Paris graduate \vplink{Buu-Hoi+Nguyen-Phuc}{Buu-Hoi Nguyen-Phuc} to early American doctorates like \vplink{Xuong+Nguyen-Huu}{Xuong Nguyen-Huu} (1962, UC Berkeley) to today's newest assistant professors.
\item \textbf{Exploring institutional and geographic patterns.} The aggregated view (Section~\ref{sec:results}) allows visitors to see how diaspora faculty are distributed across disciplines, institutions, and countries.
\end{itemize}

It is equally important to clarify what VietProfs is \emph{not}.
It is not a ranking system, and it will never rank scholars or institutions.
Nor is it a complete census of every Vietnamese-heritage academic worldwide: coverage inevitably reflects public web visibility and search coverage (Section~\ref{sec:limitations}).
The roster represents the scholars who have been identified and rigorously verified, not the absolute total population.

\section{Who Is Included: Eligibility}
\label{sec:eligibility}
A directory is only as useful as its inclusion rule is clear, and only as trustworthy as that rule is applied consistently.
This section outlines who is eligible to appear in VietProfs and why.

\subsection{The core requirement}

\begin{table}[t]
\centering
\caption{Accepted appointment tracks and their intent (\codelinkd{ROSTER\_MAINTENANCE.md}{ROSTER\_MAINTENANCE.md}).}
\label{tab:tracks}
\small
\begin{tabularx}{\linewidth}{@{}lX@{}}
\toprule
Track & Who it covers \\
\midrule
Tenure-line & Tenure-track or already-tenured faculty (Assistant, Associate, or Full Professor). \\
Teaching & Stable, full-time, continuing non-tenure-track teaching faculty, including Professor of Practice and equivalent titles, confirmed by institutional documentation. \\
Research & A stable, faculty-level or faculty-equivalent appointment at a university or eligible public/nonprofit scholarly research institute---not a postdoctoral or temporary role. \\
Clinical & A stable clinical-faculty appointment such as Clinical Professor or a documented clinical-faculty ladder---not an adjunct or temporary clinical role. \\
Academic staff & A university librarian or archivist with documented faculty status or a senior, permanent academic appointment. \\
Emeritus & A formally conferred emeritus/emerita title following an academic career, evidenced by an active emeritus listing or conferral record. \\
Deceased & Historical scholars who held an eligible faculty or permanent research appointment outside Vietnam during their career, documented with authoritative biographical records. \\
\bottomrule
\end{tabularx}
\end{table}

A person is included only if reliable evidence---preferably an official institutional page---supports two things at once: (1) a verified academic appointment at a university or eligible public/nonprofit research institute outside Vietnam (with exceptions for Emeritus and Deceased entries), and (2) that appointment falling into one of seven accepted tracks, summarized in Table~\ref{tab:tracks}.

\emph{Outside Vietnam} is a deliberate scope choice, not a value judgment on universities in Vietnam. In Vietnam, nearly all faculty are Vietnamese, and domestic university directories already list them. The diaspora, however, is scattered across hundreds of institutions globally with no centralized directory, which is the specific gap VietProfs addresses.

Each track has clear representatives in the roster: Nam-Trung Nguyen holds a Tenure-line professorship in engineering at Griffith University; \vplink{Thanh+Tran}{Thanh Tran} is a Professor in the Practice of electrical and computer engineering at Rice University under the Teaching track; \vplink{Khiem+Pham-Nguyen}{Khiem Pham-Nguyen} holds a Clinical Assistant Professorship in orthodontics at Boston University; and Xuong Nguyen-Huu holds an Emeritus appointment at UC San Diego.

\subsection{Who is not included, and why}
We deliberately exclude temporary or term-limited positions, such as adjuncts, visiting scholars, postdocs, graduate teaching assistants, and industry-only roles. Corporate research labs (including commercial AI labs) are outside our scope; non-university employers are limited to public or independent nonprofit research institutes, whose entries are explicitly labeled by institution type.

This is not a reflection of the quality of researchers in these roles---many are outstanding scholars, and some will later transition into eligible appointments. Rather, it is a practical choice about directory stability. Visiting and postdoctoral appointments usually last only one to three years. Including them would introduce massive turnover and leave the directory full of stale, outdated entries.

Similarly, a plain title of \code{Instructor} is evaluated case-by-case, as the title can designate either a stable teaching-track faculty member or a single-course temporary contract depending on the institution.
Research Assistant Professors are included only when the institution treats the position as a genuine faculty rank with a defined ladder, rather than a temporary postdoc title.
Professor of Practice and its equivalents are included under the Teaching track once their stability is confirmed, preserving the institution's official title rather than flattening it into a generic rank.

\subsection{Identity is not inferred from a name}
We use common Vietnamese given names and surnames (\code{Nguyen}, \code{Tran}, \code{Le}, \code{Pham}, \code{Vo}, \code{Vu}, \code{Bui}, \code{Do}, \code{Phan}, etc.) as initial leads to discover candidate profiles, just as a human researcher would.
However, a name is only a starting point, never proof. Once an individual holds an eligible academic appointment under Table~\ref{tab:tracks}, we do not require or request documentary proof of Vietnamese heritage or ethnicity.

This choice is both ethical and practical. Ethically, demanding personal proof of heritage from scholars to list their public professional profiles would be intrusive and inappropriate. Practically, a surname is neither necessary nor sufficient evidence: many non-Vietnamese individuals share common Vietnamese surnames, while some Vietnamese-heritage scholars do not bear Vietnamese surnames due to marriage, adoption, or name changes. The academic literature on name-based ethnicity classification confirms that surnames provide only probabilistic signals requiring independent corroboration \cite{mateos2007name,ambekar2009name}.
Finally, anyone who wishes to be removed from the directory or who has been misidentified can easily request removal or correction through our public submission form (Section~\ref{sec:submit}).

\section{Fields of Study Covered}
\label{sec:fields}
To make over a thousand records easy to browse by discipline, every person in VietProfs is classified into one of 17 broad fields (Table~\ref{tab:fieldmap}).
Rather than inventing an arbitrary taxonomy, these 17 groups are adapted from standard classifications used in U.S. federal statistical reporting. This ensures that visitors familiar with university or funding agency structures will find the categories intuitive, and allows VietProfs data to be discussed in the context of broader national statistics on doctoral education.

Two classification systems anchor this correspondence:
\begin{itemize}
\item The National Center for Education Statistics' \emph{Classification of Instructional Programs} (CIP), the standard system U.S. institutions use to report degree programs to the federal government \cite{nces_cip2020}.
\item The National Science Foundation's Survey of Earned Doctorates (SED), which groups doctorates into broad fields for annual national reporting \cite{nsf_sed}.
\end{itemize}

\begin{table}[t]
\centering
\caption{VietProfs's 17 broad fields and their nearest correspondence in the NCES CIP family structure \cite{nces_cip2020}.}
\label{tab:fieldmap}
\small
\begin{tabularx}{\linewidth}{@{}p{0.36\linewidth}X@{}}
\toprule
VietProfs field & Nearest CIP family/families \\
\midrule
Computer \& Information Sciences & 11 Computer and Information Sciences and Support Services \\
Engineering & 14 Engineering; 15 Engineering Technologies \\
Mathematics & 27 Mathematics and Statistics (pure/applied mathematics) \\
Statistics \& Data Science & 27 Mathematics and Statistics (statistics/data science) \\
Physics \& Astronomy & 40 Physical Sciences \\
Chemistry & 40 Physical Sciences \\
Biological \& Biomedical Sciences & 26 Biological and Biomedical Sciences \\
Earth \& Environmental Sciences & 40 Physical Sciences (earth/atmospheric/ocean sciences); 03 Natural Resources \\
Agricultural \& Natural Resource Sciences & 01 Agriculture, Agriculture Operations; 03 Natural Resources and Conservation \\
Health Sciences & 51 Health Professions and Related Clinical Sciences (see Table~\ref{tab:healthnih}) \\
Business \& Economics & 52 Business, Management, Marketing; 45.06 Economics \\
Social \& Behavioral Sciences & 45 Social Sciences; 42 Psychology \\
Education & 13 Education \\
Humanities & 23 English Language/Literature; 16 Foreign Languages; 54 History; 38 Philosophy \\
Law \& Public Affairs & 22 Legal Professions and Studies; 44 Public Administration and Social Service \\
Arts \& Design & 50 Visual and Performing Arts; 04 Architecture and Related Services \\
Others & 30 Multi/Interdisciplinary Studies; institution-specific fields not yet mapped \\
\bottomrule
\end{tabularx}
\end{table}

Table~\ref{tab:fieldmap} shows the mapping between VietProfs fields and CIP families.
This grouping is designed for clean browsing: several fields combine closely related CIP families (e.g., Humanities groups English, history, philosophy, and languages), while Economics is grouped with Business because faculty in these areas are frequently hired and housed together.

\begin{table}[t]
\centering
\caption{VietProfs Health Sciences subfields and their closest NIH Institute/Center counterpart \cite{nih_ics}.}
\label{tab:healthnih}
\small
\begin{tabularx}{\linewidth}{@{}lX@{}}
\toprule
VietProfs subfield & Closest federal classification \\
\midrule
Nursing & National Institute of Nursing Research (NINR) \\
Dentistry & National Institute of Dental and Craniofacial Research (NIDCR) \\
Biomedical Research & National Institute of General Medical Sciences (NIGMS); NCATS \\
Clinical Medicine & Organ- and disease-focused Institutes (e.g., NCI, NHLBI, NIDDK) \\
Public Health & National Institute of Environmental Health Sciences (NIEHS); CDC \\
Pharmacy & CIP 51.20 (Pharmacy, Pharmaceutical Sciences, and Administration) \\
Medical Education & CIP 51.12 (Medicine) \\
\bottomrule
\end{tabularx}
\end{table}

Health Sciences accounts for roughly 20\% of the entire roster (228 records, 104 Clinical appointments, and multiple endowed chairs in medicine). To make this large group navigable, VietProfs further breaks it into seven subfields (Table~\ref{tab:healthnih}), aligned with the National Institutes of Health's institute structure \cite{nih_ics}.

\section{What Information Is Shown, and Why}
\label{sec:whatshown}
Every attribute displayed on a profile card (Figure~\ref{fig:home}) serves a specific, practical purpose:
\begin{itemize}
\item \textbf{Portrait:} Photos make the directory recognizable and welcoming. For community visibility, a portrait is often the quickest way to confirm identity, particularly given how common many Vietnamese surnames are. Portraits are stored locally as optimized WebP images with recorded source provenance (\code{portraitSource}).
\item \textbf{Vietnamese Name with Diacritics:} Most English-language university websites strip diacritics or reorder Vietnamese names. We record the authentic diacritic form (\code{vietnameseName}) to preserve cultural authenticity and help people recognize names properly.
\item \textbf{Institution, Department, Rank, and Location:} These core attributes establish where each scholar works and enable faceted filtering by geographic region and department.
\item \textbf{Educational Background:} Recording degrees (undergraduate, master's, PhD, postdoc, and MD/medical training) helps students find mentors who followed similar academic paths or graduated from their alma mater.
\item \textbf{Curated Honors:} We highlight major, field-wide distinctions (such as academy memberships or early-career awards) to showcase scholarly impact while avoiding routine CV clutter (Section~\ref{sec:honors}).
\item \textbf{Research Areas:} Short keyword tags support topic-based searching beyond high-level department names.
\item \textbf{Web and Profile Links:} Direct links to university directory pages, Google Scholar, and personal/lab homepages let visitors verify information and explore publications.
\item \textbf{Verification Timestamp (\code{lastUpdatedAt}):} Showing when a profile was last verified against the open web gives visitors a transparent measure of data freshness.
\end{itemize}

\section{Data Model and Curated Attributes}
\label{sec:model}
The public source of truth is \codelink{public/data.json}, loaded and queried entirely client-side. Table~\ref{tab:schema} lists the record schema implemented by \code{RosterEntry} in \codelink{src/data.ts}.
A record contains an immutable \code{vp-} identifier, name, diacritic spelling, current profile URL, university, non-university \code{institutionType}, city, state, country, department, research areas, rank, appointment track, optional education across five credential types, curated honors, optional links, portrait provenance, and \code{lastUpdatedAt}.
The separate \codelink{maintenance/verification.json} ledger records the time of completed full reviews, keyed by canonical name, allowing maintenance state to travel with the repository without inflating the client bundle.

\begin{table}[t]
\centering
\caption{Public roster schema (\code{RosterEntry}, \code{src/data.ts}).}
\label{tab:schema}
\small
\begin{tabularx}{\linewidth}{@{}l>{\raggedright\arraybackslash}X@{}}
\toprule
Group & Stored Attributes \\
\midrule
Required & \code{id}, \code{name}, \code{university}, \code{department}, \code{rank}, \code{track}, \code{profileUrl}, \code{lastUpdatedAt}, \code{researchAreas[]} \\
Location & \code{city}, \code{state}, \code{country} \\
Identity & \code{vietnameseName} (authoritative diacritic spelling) \\
Institution & \code{institutionType} (required for eligible non-university research institutes) \\
Education & \code{undergradInstitution}/\code{Year}/\code{Major}, \code{msInstitution}/\code{Year}/\code{Major}, \code{phdInstitution}/\code{Year}/\code{Major}, \code{postdocInstitution}/\code{Year}, \code{mdInstitution}/\code{Year}, \code{otherDegrees[]} \\
Honors & \code{honors[]}: \code{name}, \code{organization}, \code{category}, \code{year}, \code{source} \\
Links & \code{websiteUrl} (personal/lab page), \code{scholarUrl} (Google Scholar) \\
Portrait & \code{portrait} (local WebP path), \code{portraitSource} (provenance URL) \\
\bottomrule
\end{tabularx}
\end{table}

\subsection{Curated Honors and Awards}
\label{sec:honors}
The \code{honors} field is restricted to substantial distinctions, avoiding routine CV padding. Table~\ref{tab:honors} lists the five accepted categories. Both the controller's proposal validator and \codelink{scripts/validate-data.ts}{validate-data.ts} enforce that every stored honor declares one of these categories, together with a name, granting organization, HTTPS provenance URL, and valid calendar year.

\begin{table}[t]
\centering
\caption{Accepted honor categories (\codelinkd{ROSTER\_MAINTENANCE.md}{ROSTER\_MAINTENANCE.md}).}
\label{tab:honors}
\small
\begin{tabularx}{\linewidth}{@{}lX@{}}
\toprule
Category & Qualifying Criteria \\
\midrule
\code{academy} & Election to a recognized national academy or equivalent learned academy. \\
\code{fellow} & Fellow or honorary-member status conferred by a major disciplinary society (IEEE, ACM, AAAS). \\
\code{career\_award} & Nationally/internationally competitive early-career award (NSF CAREER, Sloan, PECASE). \\
\code{major\_award} & Field-wide medal, book prize, lifetime achievement, or test-of-time distinction. \\
\code{distinguished\_professorship} & Named endowed chair, distinguished professorship, or comparable university chair. \\
\bottomrule
\end{tabularx}
\end{table}

\section{User Interface and Interaction Model}
\label{sec:ui}
Because the directory is designed for regular community use, the web interface is built for fast, responsive browsing without server-side database latency. The entire dataset and search index are loaded directly in the client browser.

\begin{figure}[t]
\centering
\includegraphics[width=0.90\linewidth]{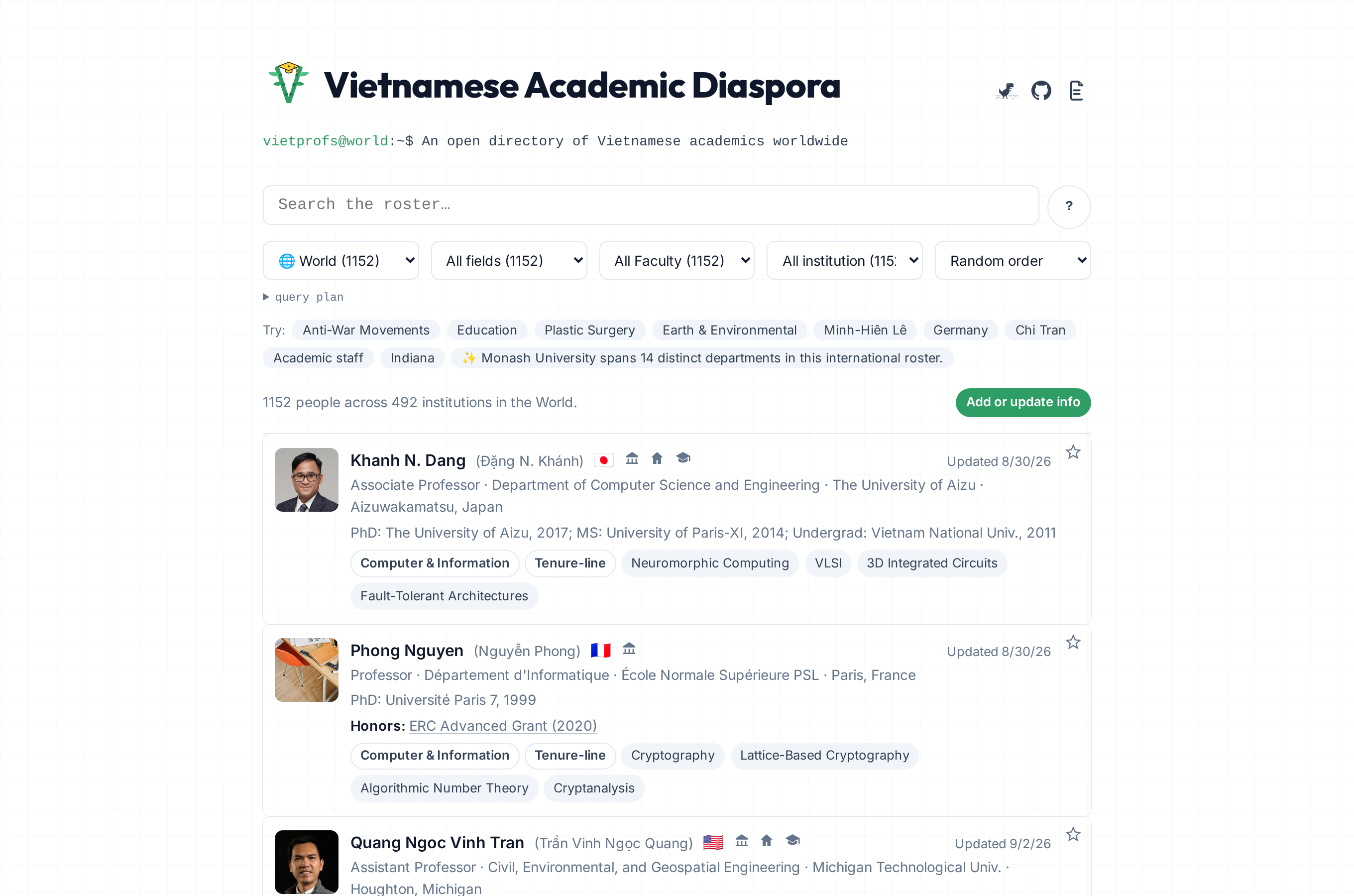}
\caption{VietProfs landing interface: search box, combinable filters (location, field, track), dynamic query chips, and profile cards with verified affiliations, education, and honors.}
\label{fig:home}
\end{figure}

Figure~\ref{fig:home} shows the landing page: a search bar with a selectable search scope, three combinable filter dropdowns (location, field of study, and appointment track), dynamic example query chips, and individual scholar cards. Each card displays the scholar's portrait, English and diacritic names, current rank, department, university, location, education history, honors, research tags, and a visible \code{lastUpdatedAt} verification date.

\subsection{Scoped Search and Diacritic Normalization}
\begin{figure}[t]
\centering
\includegraphics[width=0.90\linewidth]{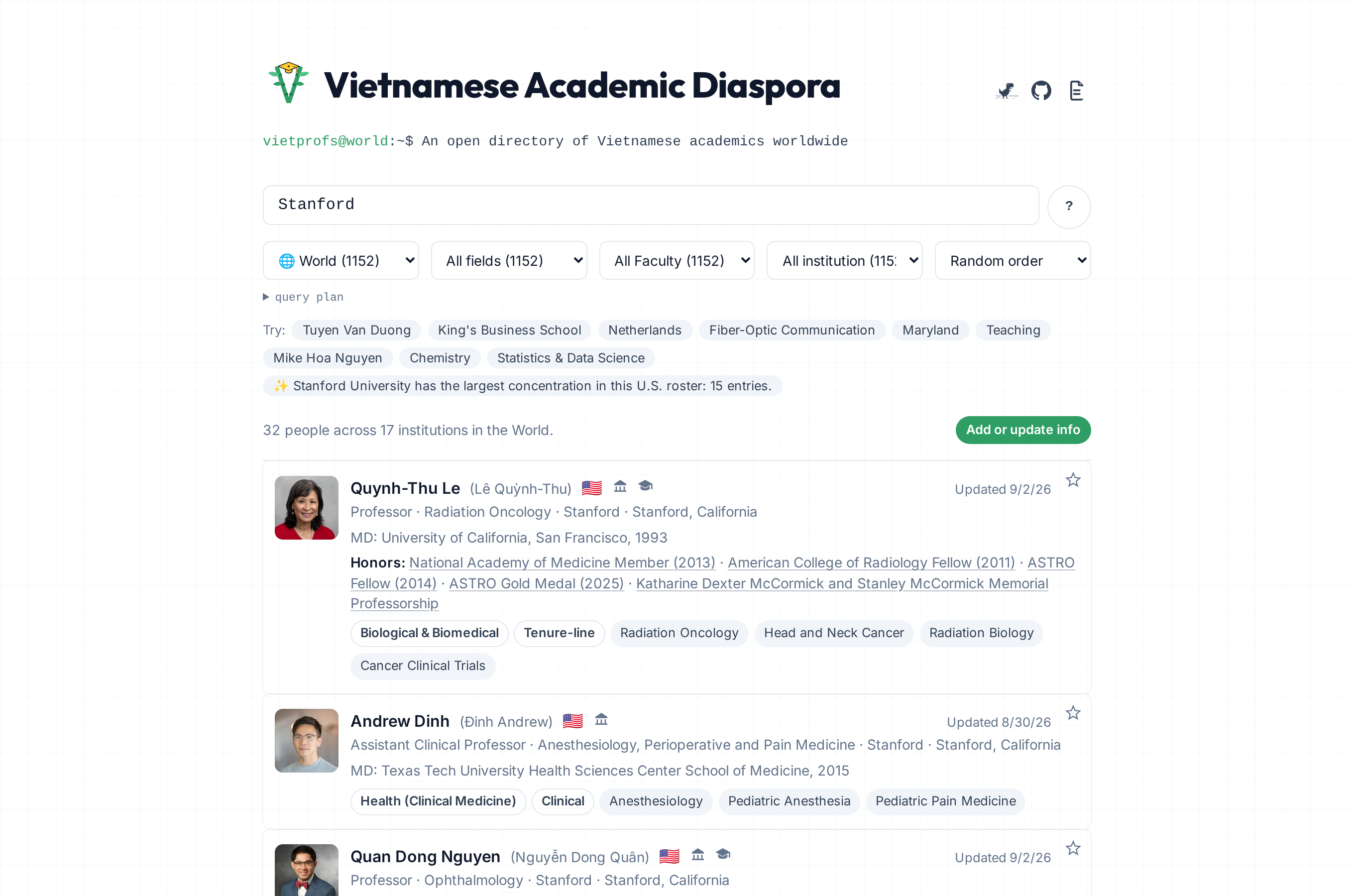}
\caption{Scoped search interface: searching within the ``University'' scope narrows the directory to scholars at a specific institution (e.g., Stanford University).}
\label{fig:search}
\end{figure}

The search bar matches across names, universities, departments, ranks, research tags, honors, and doctoral institutions. To ensure smooth searching, matching is performed against a normalized, diacritic-stripped index (\code{src/data.ts}). A user typing \code{Nguyen} will immediately match \code{Nguy\~{e}n}.
Users can also narrow search queries using a scope dropdown on the left of the search bar, restricting searches to specific attributes such as Name, Rank, Research Area, Honors, University, Department, or PhD Institution (Figure~\ref{fig:search}).
Filters are automatically synced with the URL query parameters, making any filtered view (e.g., all Computer Science faculty in California) directly bookmarkable and shareable.

\subsection{Leaderboards, Geographic Map, and Dynamic Facts}
Directly below the directory listing, the interface computes dynamic leaderboards for ``Top Faculty Hubs'' and ``Top PhD Alma Maters'' for the currently filtered subset. Each leaderboard entry is itself clickable, allowing users to drill down into specific institutions with a single click.

\begin{figure}[t]
\centering
\includegraphics[width=0.90\linewidth]{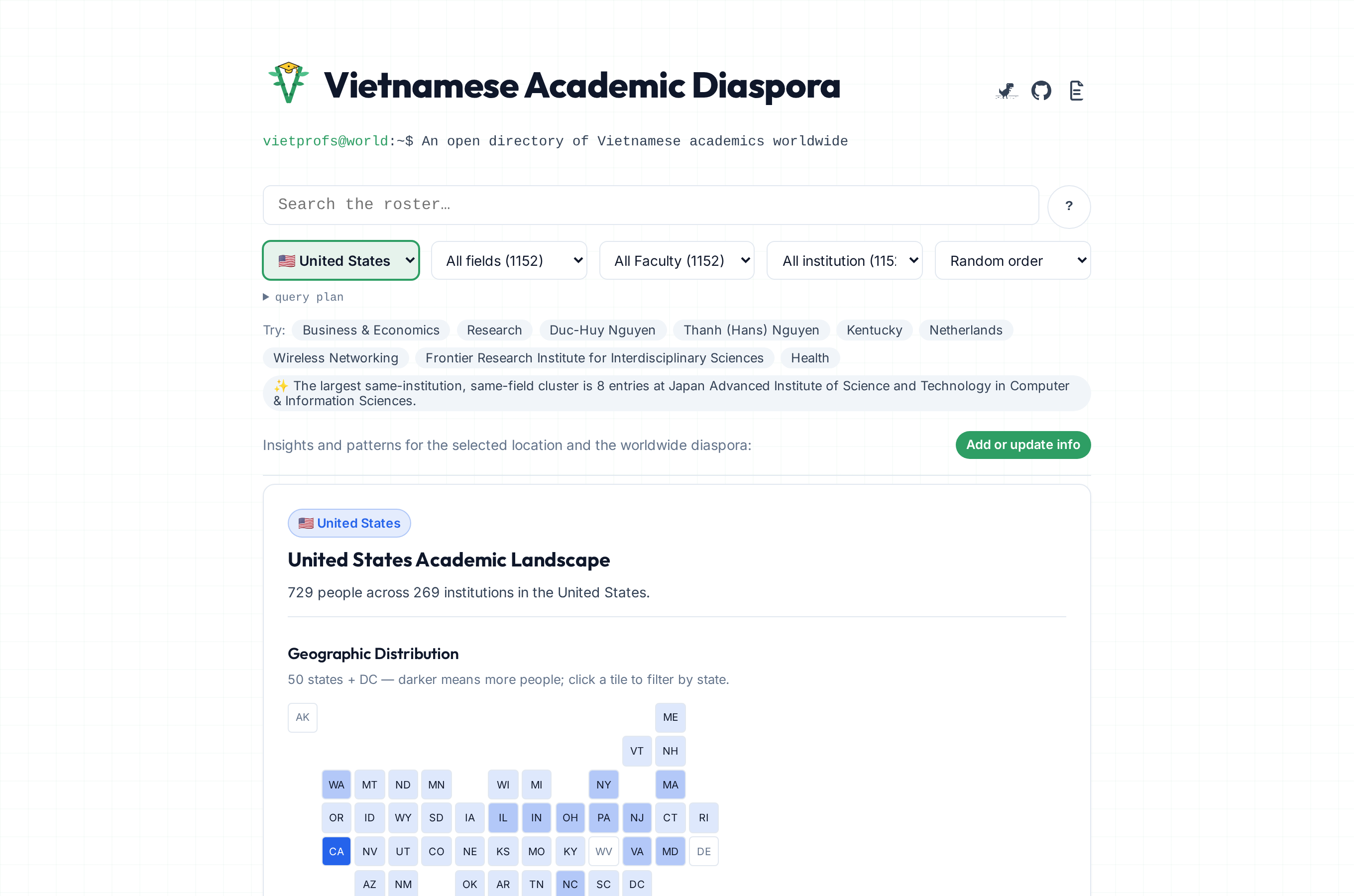}
\caption{Dynamic facts view: an interactive choropleth map of U.S. records and real-time aggregate statistics computed directly from the current snapshot.}
\label{fig:interesting}
\end{figure}

Selecting a U.S. location or clicking ``\emph{Show me something interesting}'' reveals an interactive choropleth map of the United States (\code{src/state-grid.ts}, Figure~\ref{fig:interesting}), where tile shading corresponds to faculty counts and clicking a state immediately filters the roster to that region.

\subsection{Human Contribution and Corrections Channel}
\label{sec:submit}
\begin{figure}[t]
\centering
\includegraphics[width=0.85\linewidth]{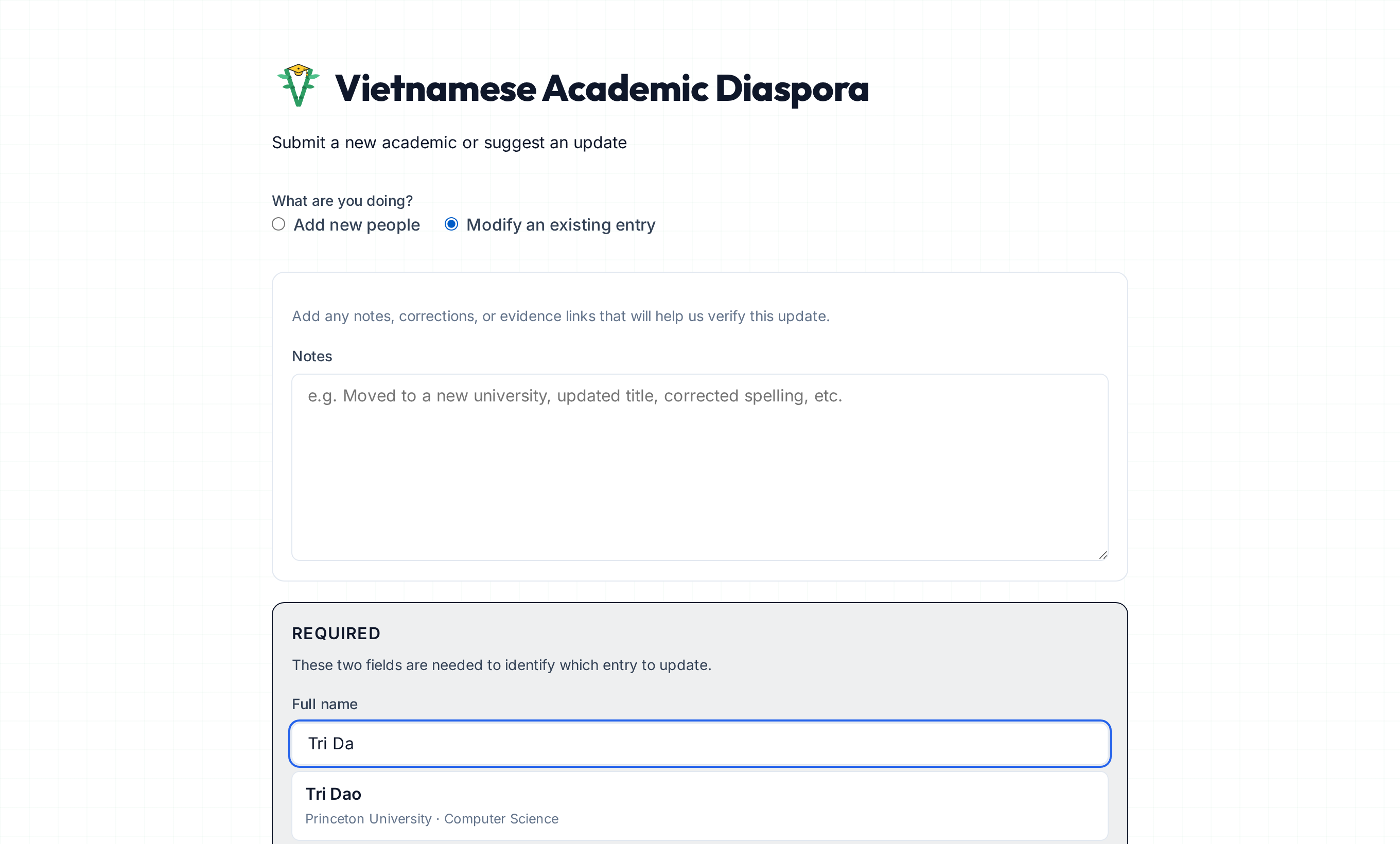}
\caption{Public submission modal: allows visitors to submit new candidates or pre-fill existing records for corrections with provenance URLs.}
\label{fig:submit}
\end{figure}

While automated workflows handle regular maintenance, community contributions are essential for discovering new scholars. Anyone can suggest a new profile or submit a correction via the public submission form at \href{https://vietprofs.roars.dev/submit.html}{\code{submit.html}} (Figure~\ref{fig:submit}).

To make submitting easy, the form requires only a scholar's name and at least one verifiable link (such as a departmental profile, personal website, or Google Scholar page). When a visitor types a name that already exists in the roster, the form automatically offers to pre-fill the existing record, allowing them to propose a correction rather than creating a duplicate entry.
All submissions are routed to a review queue where our verification pipeline and human maintainers verify the claims before merging. No user submission can directly mutate the canonical database.

\section{What the Directory Reveals}
\label{sec:results}
Beyond serving as a practical lookup tool, VietProfs provides interesting empirical visibility into the Vietnamese academic diaspora.

\subsection{Runtime Aggregation and Current Distribution}
The ``Show me something interesting'' view is computed dynamically at runtime by \code{buildFunFacts} in \codelink{src/data.ts}. Rather than hardcoding static statistics that easily become outdated, all aggregate metrics are recomputed directly from the live dataset whenever a page is rendered.

Table~\ref{tab:snapshot} summarizes the directory snapshot as of September 2026. The roster includes 1,152 verified scholars across 492 institutions in 23 countries. In the United States (729 records), California (148) and Texas (98) host the largest concentrations. The international subset (423 records) is led by Australia (95), Canada (58), the United Kingdom (57), France (46), Japan (40), and Singapore (27).
The four largest broad fields are Health Sciences (228), Business \& Economics (220), Computer \& Information Sciences (163), and Engineering (152).

\begin{table}[t]
\centering
\caption{Current VietProfs snapshot and evidence coverage (September 2026).}
\label{tab:snapshot}
\small
\begin{tabular}{lr@{\qquad}lr}
\toprule
Metric & Count & Metric & Count \\
\midrule
Total Records & 1,152 & Universities \& Institutes & 492 \\
Countries & 23 & U.S. / International & 729 / 423 \\
Broad Fields & 17 & Accepted Tracks & 7 \\
Profile URLs & 1,152 & Google Scholar URLs & 662 \\
Personal / Lab URLs & 211 & PhD Institutions & 868 \\
Honors-Bearing Records & 208 & PhD Graduation Years & 613 \\
Earliest Doctoral Year & 1939 & Latest Doctoral Year & 2026 \\
\bottomrule
\end{tabular}
\end{table}

\subsection{Historical Breadth and Multi-Generational Legacy}
Doctoral graduation dates recorded in VietProfs span 87 years (1939--2026). Early pioneers include \vplink{Buu-Hoi+Nguyen-Phuc}{Buu-Hoi Nguyen-Phuc} (1939 Paris doctorate; CNRS Director of Research) and \vplink{Xuong+Nguyen-Huu}{Xuong Nguyen-Huu} (1962 UC Berkeley PhD; UCSD), demonstrating that Vietnamese scholarly contributions abroad represent a sustained, multi-generational presence rather than a single recent wave.

Because derived observations depend on primary records, whenever a scholar relocates or an appointment track is updated, dependent statistics update automatically. VietProfs recomputes all aggregated metrics at render time, ensuring stale statistical claims are never stored.

\section{Limitations and Ethical Considerations}
\label{sec:limitations}
VietProfs is a curated public directory, not a demographic census. Discoverability depends heavily on institutional web transparency, English-language profile availability, and search effort. Consequently, directory counts should not be interpreted as demographic totals, migration paths, or indicators of institutional prestige.
Ethically, the directory processes only publicly available professional data, strictly avoiding invasive ethnicity tests, private contact information, or unverified claims.

\vspace{0.6em}
\noindent\rule{\linewidth}{0.4pt}
\begin{center}
{\Large\textbf{Part II: Building and Maintaining VietProfs: A Technical Report}}
\end{center}
\noindent\rule{\linewidth}{0.4pt}
\vspace{0.4em}

\section{Why Maintaining the Directory Is Hard}
\label{sec:hard}
Keeping a public directory accurate against the open web involves five primary engineering challenges:
\begin{enumerate}
\item \textbf{Open-World Search:} There is no central registry of Vietnamese-diaspora scholars; the population is distributed globally across thousands of departments.
\item \textbf{Heterogeneous Web Sources:} Department pages, university directories, personal CVs, and lab websites vary wildly in format, freshness, and authority.
\item \textbf{Non-Standard Academic Titles:} Titles such as ``Lecturer'' (UK/Australia vs. US) or ``Research Assistant Professor'' carry very different contractual and tenure meanings across institutions and countries.
\item \textbf{Name Disambiguation:} Vietnamese surnames are highly concentrated; verifying an individual requires joint checking of full names, institutional affiliations, and research areas.
\item \textbf{Link Rot and Site Redesigns:} Universities regularly redesign websites and migrate CMSs, causing hyperlinks to break \cite{bar-yossef2004sic}. A 404 error must be distinguished from a genuine departure or retirement.
\end{enumerate}

\section{System Architecture}
\label{sec:architecture}
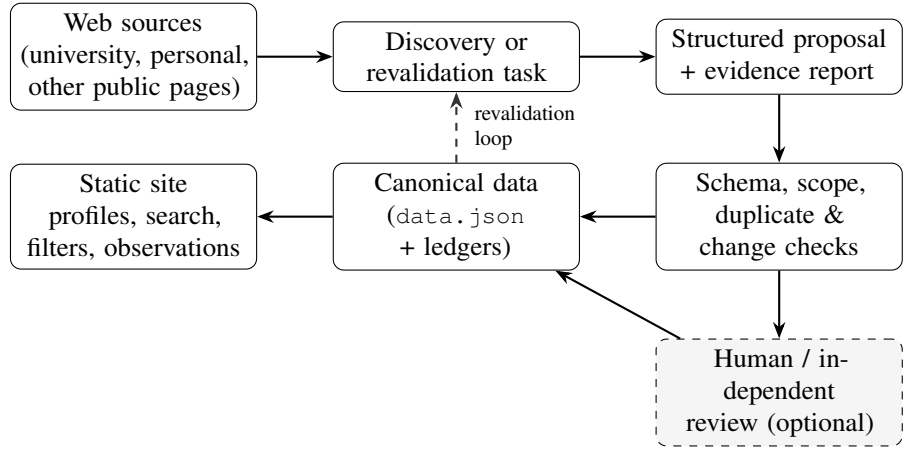
\begin{figure}[t]
\centering
\begin{tikzpicture}[
  node distance=9mm and 10mm,
  every node/.style={font=\small},
  box/.style={draw, rounded corners, align=center, minimum height=9mm, text width=30mm},
  opt/.style={draw, rounded corners, align=center, minimum height=8mm, text width=30mm, fill=gray!8, dashed},
  arr/.style={-{Stealth[length=2.2mm]}, thick},
  loop/.style={-{Stealth[length=2.2mm]}, thick, dashed, gray!40!black}
]
\node[box] (web) {Web sources\\(university, personal, other public pages)};
\node[box, right=of web] (task) {Discovery or\\revalidation task};
\node[box, right=of task] (proposal) {Structured proposal\\+ evidence report};

\node[box, below=of proposal] (gate) {Schema, scope,\\duplicate \& change checks};
\node[box, left=of gate] (canonical) {Canonical data\\(\texttt{\footnotesize data.json} + ledgers)};
\node[box, left=of canonical] (site) {Static site\\profiles, search, filters, observations};

\node[opt, below=of gate] (human) {Human / independent\\review (optional)};

\draw[arr] (web) -- (task);
\draw[arr] (task) -- (proposal);
\draw[arr] (proposal) -- (gate);
\draw[arr] (gate) -- (canonical);
\draw[arr] (canonical) -- (site);
\draw[arr] (gate) -- (human);
\draw[arr] (human) -- (canonical);
\draw[loop] (canonical.north) -- node[midway, right=1mm, font=\scriptsize, black, align=left] {revalidation\\loop} (task.south);
\end{tikzpicture}
\caption{VietProfs architecture: bounded web research feeds structured proposals through deterministic validation gates into the canonical repository; periodic revalidation completes the loop.}
\label{fig:arch}
\end{figure}

Figure~\ref{fig:arch} illustrates the system architecture. Candidate discovery and periodic revalidation feed structured evidence to LLM agents. Proposed patches must pass deterministic schema, track, URL, and target-scope validation before the controller mutates the canonical JSON repository. Table~\ref{tab:pipeline} lists the key implementation subsystems.

\begin{table}[t]
\centering
\caption{Subsystem inventory and automation boundaries.}
\label{tab:pipeline}
\small
\begin{tabularx}{\linewidth}{@{}>{\raggedright\arraybackslash}p{0.18\linewidth}>{\raggedright\arraybackslash}p{0.39\linewidth}X@{}}
\toprule
Subsystem & Implementation & Operational Boundary \\
\midrule
Discovery queries & \codelinkd{scripts/faculty-discovery-queries.ts}{faculty-discovery-queries.ts} & Deterministic query generator; candidate research is agent/human. \\
Maintenance engine & \codelinkd{scripts/maintain-roster.ts}{maintain-roster.ts} & Unattended controller; invokes LLM agent (Claude/Codex) via CLI. \\
Proposal validation & Controller + \codelinkd{scripts/validate-data.ts}{validate-data.ts} & Deterministic schema, track, URL, and target-scope enforcement. \\
Canonical data & \codelinkd{public/data.json}{data.json} & Immutable-ID JSON repository; modified only by validated patches. \\
Verification ledger & \codelinkd{maintenance/verification.json}{verification.json} & Timestamped ledger tracking when each record was last verified. \\
Static profiles & \codelinkd{scripts/generate-profile-pages.ts}{generate-profile-pages.ts} & Deterministic generator creating individual static HTML pages. \\
Client app & \codelinkd{src/data.ts}{data.ts}, \codelinkd{src/main.ts}{main.ts} & In-browser search, indexing, filtering, and dynamic fact aggregation. \\
\bottomrule
\end{tabularx}
\end{table}

\subsection{Controller Execution and Proposal Validation}
The maintenance controller (\codelink{scripts/maintain-roster.ts}) manages unattended execution batches. It selects stale or unverified records from the verification ledger (\code{--stale-days}, default 365), invokes the research agent with tool access restricted to read-only browsing (\code{WebSearch}, \code{WebFetch}), and receives a structured JSON proposal. Table~\ref{tab:cli} summarizes the controller CLI options.

\begin{table}[t]
\centering
\caption{Maintenance controller command-line options (\codelink{scripts/maintain-roster.ts}).}
\label{tab:cli}
\small
\begin{tabularx}{\linewidth}{@{}lX@{}}
\toprule
Flag & Operational Behavior \\
\midrule
\code{--limit N} & Process up to $N$ entries in a single batch (default 40). \\
\code{--total N} & Process up to $N$ entries across consecutive batches, committing after each. \\
\code{--all} & Full-sweep mode: process all stale entries across the entire directory. \\
\code{--stale-days N} & Target records whose last full verification is older than $N$ days (default 365). \\
\code{--name QUERY} & Direct target mode: resolves and processes a specific scholar or field slice. \\
\code{--dry-run} & Performs research and validation without writing changes to Git or disk. \\
\code{--codex-review} & Requires an independent second-agent review verdict before applying changes. \\
\code{--agent A} & Selects the primary research model (\code{claude} default, or \code{codex}). \\
\bottomrule
\end{tabularx}
\end{table}

Before applying any change, the controller enforces strict deterministic invariants:
\begin{itemize}
\item \textbf{Schema Conformance:} Mandatory fields must be present; URLs must be valid HTTP(S); appointment tracks must match the seven accepted values.
\item \textbf{Target Scoping:} Proposals cannot alter records other than the targeted scholar or reorder the database.
\item \textbf{Provenance Enforcement:} Stored honors must include valid HTTPS sources, accepted categories, and non-future years.
\end{itemize}

\subsection{Independent Review and Resumable Checkpoints}
When \code{--codex-review} is active, a second model independently reviews proposed patches against web evidence. If rejected, the controller permits up to two bounded revision attempts (\code{MAX\_PROPOSAL\_REVISIONS = 2}). If discrepancies remain, the case is deferred rather than guessed.
Working state is written to disk (\code{\textasciitilde/.local/state/vietprofs-maintenance/}), allowing long maintenance runs to resume seamlessly across process restarts or rate limits.

\section{Discovery as Partitioned Open-World Search}
Unbounded search queries like ``find all Vietnamese faculty worldwide'' are intractable. VietProfs structures candidate discovery by partitioning open search into bounded institution, department, and domain slices using \codelink{scripts/faculty-discovery-queries.ts}.

Queries combine institutional keywords, faculty directory paths, and Vietnamese surname/name tokens, paired with negative filter keywords (\code{-student -postdoctoral}) to minimize noise. Discovered leads are checked against the canonical roster before undergoing full evidence verification.

\section{Evidence and Verification}
\label{sec:evidence}
Candidate discovery is strictly separated from verification. An official university faculty page remains the primary standard of evidence. Personal lab websites, Google Scholar profiles, and disciplinary award announcements provide corroborating details for education and honors.
Under the operating principle \emph{AI proposes, evidence decides} \cite{amershi2014power}, model outputs are treated solely as structured hypotheses until validated against web provenance.

\section{Continuous Revalidation and Self-Maintenance}
\label{sec:maintenance}
\begin{figure}[t]
\centering
\begin{tikzpicture}[node distance=8mm and 8mm, every node/.style={font=\small}, box/.style={draw, rounded corners, align=center, minimum height=8mm, text width=25mm}, arr/.style={->, thick}]
\node[box] (unknown) {Unknown\\external person};
\node[box, right=of unknown] (candidate) {Candidate\\with leads};
\node[box, right=of candidate] (verified) {Verified\\record};
\node[box, right=of verified] (published) {Published\\snapshot};
\node[box, below=of published] (recheck) {Periodic\\revalidation};
\node[box, left=of recheck] (changed) {Unchanged,\\updated, or ambiguous};
\draw[arr] (unknown)--(candidate)--(verified)--(published)--(recheck)--(changed);
\draw[arr] (changed.north) |- (published.south);
\end{tikzpicture}
\caption{Continuous maintenance lifecycle: newly discovered candidates and stale roster entries undergo evidence verification before canonical commitment.}
\label{fig:lifecycle}
\end{figure}
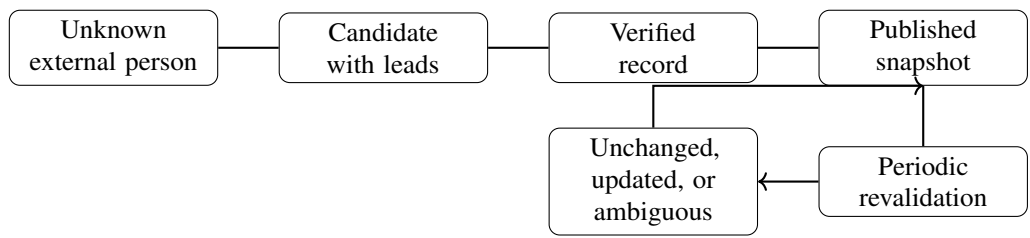

Figure~\ref{fig:lifecycle} depicts the two continuous loops governing directory maintenance: expansion and revalidation.

The periodic revalidation workflow executes six sequential checks:
\begin{enumerate}
\item \textbf{Profile Liveness:} Verify profile URL accessibility and identity matching; detect departures or moves.
\item \textbf{URL Role Disambiguation:} Ensure institutional directory pages and personal/lab homepages are correctly assigned.
\item \textbf{Scholar Verification:} Match Google Scholar profiles using publication titles and departmental affiliations.
\item \textbf{Appointment \& Rank Audit:} Detect promotions, title updates, and track reclassifications.
\item \textbf{Education \& Honors Backfill:} Extract verified degree institutions, graduation years, and curated major awards.
\item \textbf{Ledger Update:} Record timestamped verification in \codelink{maintenance/verification.json}.
\end{enumerate}

\section{Testing and Validation}
\label{sec:testing}
Automated maintenance is protected by a multi-layered test suite executed on every commit and pull request:
\begin{itemize}
\item \textbf{Data Invariants (\codelink{scripts/validate-data.ts}):} Re-verifies all schema rules, URL schemes, portrait existence on disk, unique IDs, and synchronization with the verification ledger.
\item \textbf{Unit Tests (\codelink{test/data.test.ts}, \codelink{test/maintenance-controller.test.ts}):} Tests field-mapping heuristics, search indexing, diacritic handling, dynamic fact generation, and proposal analysis.
\item \textbf{Route Integrity (\codelink{test/profile-pages.test.ts}):} Verifies static profile page generation and immutable URL stability.
\item \textbf{End-to-End Smoke Tests (\codelink{test/browser-smoke.test.ts}):} Headless Playwright browser tests verifying search interactions, filter combinations, and submission flows on the built site.
\end{itemize}

\section{Operational Experience and Evaluation}
Table~\ref{tab:cases} documents representative maintenance events from the Git audit log. These cases illustrate how automated proposals, deterministic gates, and human oversight interact to resolve duplicates, correct appointment tracks, update moved affiliations, and record negative search results.

\begin{table}[t]
\centering
\caption{Representative maintenance events in the public Git history.}
\label{tab:cases}
\small
\begin{tabularx}{\linewidth}{@{}p{0.18\linewidth}p{0.26\linewidth}X@{}}
\toprule
Case & Commit & Action and Operational Rationale \\
\midrule
Duplicate & \commitlink{2aa15c0} & Merged duplicate entries for Tam V. Nguyen at University of Dayton. \\
Policy removal & \commitlink{9a4a171} & Removed temporary non-faculty role under inclusion criteria. \\
Track correction & \commitlink{ba71beb} & Reclassified clinical dental faculty from Teaching to Clinical track. \\
Affiliation move & \commitlink{c1f4380} & Updated moved institution and profile URL for Nguyen-Truc-Dao Nguyen. \\
Name cleanup & \commitlink{c0c0603} & Standardized diacritic spelling and resolved redundant name suffixes. \\
Negative sweep & \commitlink{9b31b95} & Recorded completed regional audit with zero false-positive additions. \\
\bottomrule
\end{tabularx}
\end{table}

\section{What Does Not Work}
\label{sec:whatdoesnotwork}
Several intuitive automation shortcuts failed in practice and had to be abandoned:
\begin{itemize}
\item \textbf{Unconstrained LLM Querying:} Asking models to list faculty from memory without external search tools produces convincing hallucinations: non-existent appointments, fabricated departments, and wrong institutions \cite{ji2023survey}.
\item \textbf{Surname-Only Filtering:} Inferring diaspora status purely from common surnames creates high false-positive rates (e.g., non-Vietnamese scholars sharing similar romanized names) while missing Vietnamese scholars with non-Vietnamese surnames \cite{mateos2007name}.
\item \textbf{Treating 404s as Departures:} Campus websites break links constantly during CMS overhauls. Deleting a profile solely because a link returned a 404 resulted in erroneously removing active faculty.
\item \textbf{Direct Model Writes to the Database:} Allowing an LLM to directly write or format JSON records frequently corrupted schema structures, reordered entries, or stripped existing fields.
\end{itemize}

\section{Practical Engineering Lessons}
Our experience maintaining VietProfs yields ten practical takeaways for agent-maintained datasets:
\begin{enumerate}
\item \textbf{AI Proposes, Determinism Decides:} Restrict LLMs to information retrieval and proposal generation; enforce schema and business invariants using deterministic code.
\item \textbf{Decouple Candidate Leads from Canonical State:} Route external suggestions and unverified leads into a staging pipeline rather than directly mutating production data.
\item \textbf{Target-Scoped Edits:} Force agent proposals to touch only the targeted record, preventing batch-wide corruption.
\item \textbf{Compute Aggregate Facts at Runtime:} Never store precomputed summary statistics that can silently become stale when underlying records change.
\item \textbf{Require Corroboration for Removals:} Never remove an entry based on a broken link alone; demand secondary evidence of departure or retirement.
\item \textbf{Track a Two-Tiered Verification Ledger:} Distinguish between the date a record was modified and the date it was audited and confirmed unchanged.
\item \textbf{Curate Domain Honors Strictly:} Avoid uncurated CV scraping; restrict honors to standardized, comparable award tiers.
\item \textbf{Partition Open-World Search:} Structure broad searches into narrow, repeatable institution-by-field search slices.
\item \textbf{Use Multi-Agent Cross-Checking:} Require secondary model review and bounded retry cycles for ambiguous claims.
\item \textbf{Guard Production with End-to-End Tests:} Protect deployment with automated linting, schema validation, and headless browser tests.
\end{enumerate}

\section{Related Work}
\label{sec:related}
VietProfs connects to several areas of related work:
\begin{itemize}
\item \textbf{Entity Resolution and Web Extraction:} Traditional entity resolution focuses on blocking and deduplication across relational databases \cite{christophides2019er,elmagarmid2007duplicate}. Recent efforts explore LLM prompting for entity matching \cite{peeters2023llm,li2024er}. VietProfs applies these insights specifically to scholarly appointment verification.
\item \textbf{Scholarly Knowledge Graphs:} Platforms like OpenAlex \cite{priem2022openalex}, DBLP, and ORCID index publications and citation networks. VietProfs focuses on the orthogonal problem of verifying active university faculty appointments, tracks, and diaspora community visibility.
\item \textbf{Federal Educational Taxonomies:} VietProfs grounds its browsing structure in official federal classifications (NCES CIP \cite{nces_cip2020}, NSF SED \cite{nsf_sed}, and NIH ICs \cite{nih_ics}), allowing direct comparison with national doctoral statistics.
\end{itemize}

\section{Conclusion and Future Work}
\label{sec:conclusion}
VietProfs demonstrates that a specialized, evidence-backed public directory can be built and maintained over time without requiring an army of human curators.
By pairing automated agent research with strict deterministic validation gates, the system keeps over a thousand academic records accurate against a constantly changing web.
For the first time, prospective students can find mentors who share their background in seconds; researchers can discover colleagues across institutions and borders; and the Vietnamese academic diaspora has an authoritative record of its own reach across generations.

Looking ahead, we plan to expand VietProfs along several practical directions:
\begin{itemize}
\item \textbf{Academic Genealogy:} Mapping mentor-student lineages across generations of Vietnamese scholars to visualize how academic mentorship has evolved.
\item \textbf{Proximity and Mentorship Matching:} Helping students connect with nearby faculty mentors across institutions for informal advising and career guidance.
\item \textbf{Collaboration Bridges with Vietnam:} Assisting domestic universities and research institutes in Vietnam to connect with diaspora faculty for guest lectures, joint grants, and sabbatical visits.
\item \textbf{Self-Service Verification:} Allowing listed scholars to claim, verify, and update their own profiles directly through institutional email verification, while maintaining deterministic schema invariants.
\end{itemize}

Throughout all these extensions, our core design philosophy remains unchanged: \emph{AI proposes, evidence decides}.
By anchoring automated tools to verifiable public evidence, we can keep community resources reliable, open, and lasting.

\clearpage
\bibliographystyle{plain}
\bibliography{references}

\end{document}